\documentclass[superscriptaddress,secnumarabic,
amssymb,amsmath,nobibnotes,aps,prd,showkeys,showpacs,nofootinbib]{revtex4}%
\usepackage{graphicx}
\usepackage{epsf}
\usepackage{bm}
\usepackage{amsmath}
\usepackage{amsfonts}
\usepackage{amssymb}
\usepackage{epstopdf}
\usepackage{natbib}%
\providecommand{\U}[1]{\protect\rule{.1in}{.1in}}
\newcommand{\be}{\begin{equation}}
\newcommand{\ee}{\end{equation}}

\newcommand{\mincir}{\raise
-3.truept\hbox{\rlap{\hbox{$\sim$}}\raise4.truept\hbox{$<$}\ }}
\newcommand{\magcir}{\raise
-3.truept\hbox{\rlap{\hbox{$\sim$}}\raise4.truept\hbox{$>$}\ }}

\usepackage{color}
\begin{document}
\title{Perfect Fluid Cosmological Universes: One equation of state and the most general solution}
\author{Anadijiban Das}
\email{das@sfu.ca}
\affiliation{Department of Mathematics, Simon Fraser University, Burnaby, British Columbia, V5A 1S6, Canada}
\author{Asit Banerjee}
\email{asitban@yahoo.co.in}
\affiliation{Department of Physics, Jadavpur University, West Bengal, Kolkata $-$ 700032, India}
\author{Subenoy Chakraborty}
\email{schakraborty@math.jdvu.ac.in}
\affiliation{Department of Mathematics, Jadavpur University, West Bengal, Kolkata $-$ 700032, India}
\author{Supriya Pan}
\email{span@iiserkol.ac.in}
\affiliation{Department of Physical Sciences, Indian Institute of Science Education and Research,
Kolkata, Mohanpur $-$ 741246, West Bengal, India}
\begin{abstract}
Considering a homogeneous and isotropic universe characterized by the Friedmann-Lema\^itre-Robertson-Walker (FLRW) line element, in this work, we have prescribed a general formalism for the cosmological
solutions when the equation of state of the cosmic substance follows a general structure $\phi (p, \rho) = 0$, where $p$, $\rho$ are respectively the pressure and the energy density of the cosmic substance. Using the general formalism we recover some well known solutions, namely, when the cosmic substance obeys the linear equation of state, a Chaplygin type equation of state, or a nonlinear equation of state. Thus, the current work offers a new technique to solve the cosmological solutions without any prior relation between $p$ and $\rho$. 

\end{abstract}
\maketitle

\section{Introduction}
\label{intro}

During the present century, a large number of observational results point to the overall regularities, which are global rather
than local. The simplest and the most elegant assumption is that, our universe is homogeneous and isotropic on the large scales of our universe. Geometrically,
the metric in this case can be expressed in the Friedmann-Lema\^{i}tre-Robertson-Walker (FLRW) form and the dynamics of the universe is governed by the Einstein's field equations $G_{\mu \nu}= 8 \pi G \, T_{\mu \nu}$, where the left hand side carries the information of the geometrical sector, $T_{\mu \nu}$ is the energy-momentum part of the matter distribution and $8 \pi G$ is the Einstein's gravitational constant. 
The right hand side
of the field equations, i.e. $T_{\mu \nu}$ contain the cosmological fluid consisting of dust, radiation and the so called vacuum energy. Various forms
of dark energy may also appear, where the violation of the strong energy condition takes place that is effectively $(\rho+ 3 p)< 0$ in which $\rho$, $p$ respectively stand for the energy density and the pressure of the dark energy. It is an intriguing fact that dark energy has become a fundamental problem in modern cosmology.

If the background universe is described by the FLRW line element, then the dynamics of the universe can in principle be determined once a relation between the energy density and the pressure of the cosmic fluid is prescribed. This is the most common and viable approach to investigate the cosmic evolution. While sometimes, the expansion rate of the FLRW universe with different functional forms might be assumed. However, although both the approaches are phenomenological and perhaps be equally valid  but nevetheless, here we are interested in the first proposal. Toward this direction,
many attempts have already been done starting from the simple relation $p= w \rho$, a barotropic equation of state\footnote{We note that here we call $w$ as the equation of state parameter.} to some complex one, but with an explicit relation between $p$ and $\rho$, of course. However, one may think of what happens if one does not prescribe any equation of state but tries to find the cosmological solutions, maybe in an implicit way. Of course this could be very interesting since the cosmological solutions for other specific relations can be easily recovered. This is the motivation of the paper in which we seek for the cosmological solutions when we do not prescribe any particular equation of state.

In this paper, we have presented the most general solution of the cosmological equations with the scale factor $a(t)$ expressed
in an implicit form. This is possible when we assume an equation of state connecting pressure and density. The most important
feature of the present work is that, it is a method to obtain the solution for the scale factor for any fluid characterized
by an arbitrary equation of state. We have given examples of linear equation of state in a perfect fluid, the Chaplygin gas
representing dark energy with a special equation of state, and a general equation of state in nonlinear form.
The well known solutions for radiation universe already existing in
literature have been derived. One must note that, in the present context, the exact solution can not always be expressed in the
explicit form. In our opinion, beyond all the known homogeneous and isotropic solutions, it is important to explore many other
such solutions using the present general approach.

The manuscript is organized in the following way. In section \ref{section2}, we have established the general formalism to find out the cosmological solutions. In the next sections we have shown how one can recover the corresponding cosmological solutions for certain specific equations of state using the general formalism. In section \ref{sec:lineos}, \ref{sec:Chaplygin}, and \ref{section:nonlinear}, we have taken a linear equation of state, a generalized Chaplygin fluid, and a general class of nonlinear equation of state and presented their cosmological solutions. Finally, section \ref{conclu} concludes our work. 

\section{Notations and Field equations}
\label{section2}

In this section, physical units are so chosen that \cite{Das2012}, $G= c= 1$, $k= 8 \pi$. Roman indices take values $\{1,~2,~3,~4\}$, Greek
indices are $\{1,~2,~3\}$. Spatial coordinates are denoted by $\{x^1,~x^2,~x^3\}$, while the time co-ordinate is specified by
$x^4 \equiv t$. Einstein's field equations (involving the cosmological constant $\Lambda$) for a perfect fluid source are
furnished by

\begin{eqnarray}
G^i_j+ k \left[(\rho+ p) u^i u_j+ p \delta^i_j \right]- \Lambda \delta^i_j &=& 0,\label{Einstein-equation}
\end{eqnarray}
where $G^i_j$'s are the Einstein tensor, $u^i$ is the fluid velocity four vector with

\begin{eqnarray}
u^i u_i= -1~~~(u^4> 0).\label{fluid-velocity}
\end{eqnarray}
The Friedmann-Lema\^{i}tre-Robertson-Walker metric for a homogeneous isotropic universe is provided by

\begin{eqnarray}
ds^2&=& a^2 (t) \left[\frac{dr^2}{1-k_0 r^2}+ r^2 (d \theta^2+ \sin^2 \theta d \phi^2)\right]- dt^2,\label{flrw1}
\end{eqnarray}
or, in a different form:

\begin{eqnarray}
ds^2&=& \frac{a^2 (t)}{\left(1+ k_0 r^2/4\right)^2} \Bigl[dr^2+ r^2 d \theta^2+ r^2 \sin^2 \theta d \phi^2\Bigr]- dt^2,\label{flrw2}
\end{eqnarray}
Here, $k_0= 0$, $1$, $-1$, indicating the spatial curvature constants. Now,
the field equations are given by \cite{RBB1992}

\begin{eqnarray}
\frac{2 \ddot{a}}{a}+ \frac{\dot{a}^2+ k_0}{a^2}&=& - k p+ \Lambda,\label{friedmann1}
\end{eqnarray}
and
\begin{eqnarray}
3 \left(\frac{\dot{a}^2+ k_0}{a^2}\right)&=&k \rho+ \Lambda.\label{friedmann2}
\end{eqnarray}
We have used above the co-moving system of coordinates

\begin{eqnarray}
u^\alpha= 0,~~~\mbox{and}~~~u^4= 1.\label{fluid-velocity2}
\end{eqnarray}
The energy conservation equation following from the field equations (\ref{friedmann1})
and (\ref{friedmann2}) by

\begin{eqnarray}
\dot{\rho}+ 3 \frac{\dot{a}}{a} (\rho+ p)&=& 0~.\label{conservation}
\end{eqnarray}
Here, the number of unknown functions is 3, for example $a(t)$, $\rho (t)$ and $p(t)$. The number of
independent equations is 2, for example (\ref{friedmann1}) and (\ref{friedmann2}). The equation
(\ref{conservation}) is not any independent equation since it follows from the field
equations (\ref{friedmann1}) and (\ref{friedmann2}). We can, therefore, render the system determinate
by imposing one equation of state:

\begin{eqnarray}
S (\rho, p)&=& 0~.\label{eos}
\end{eqnarray}
It is obvious that,

\begin{eqnarray}
\left[\frac{\partial S (\rho, p)}{\partial \rho}\right]^2+ \left[\frac{\partial S (\rho, p)}{\partial p}\right]^2 &>& 0~.\label{condition1}
\end{eqnarray}
Therefore, either $\frac{\partial S (\rho, p)}{\partial \rho} \neq 0$, or $\frac{\partial S (\rho, p)}{\partial p} \neq 0$, or both
$\frac{\partial S (\rho, p)}{\partial \rho}$ and $\frac{\partial S (\rho, p)}{\partial p}$ are nonzero. Let us assume that $\frac{\partial S (\rho, p)}{\partial p} \neq 0$.

Then by the implicit function theorem \cite{MT1988}, there exists a function $p (\rho)$, such that, the pressure

\begin{eqnarray}
p&=& p (\rho),\label{condition2}
\end{eqnarray}
and

\begin{eqnarray}
\frac{dp}{d \rho}&=& -\left[\frac{\partial S (\rho, p)}{\partial \rho}/\frac{\partial S (\rho, p)}{\partial p}\right].\label{condition3}
\end{eqnarray}
We would now explore the system of field equations in the radiation era and in matter dominated era. Therefore, we assume the following physically
appropriate inequalities $a> 0$, $\rho> 0$, and $p> 0$.\\

We digress here slightly to deduce some consequences of the field equations (\ref{friedmann1}) and (\ref{friedmann2}). We get clearly
from these equation

\begin{eqnarray}
\ddot{a}&=& -\frac{k}{6} (\rho+ p) a+ \frac{\Lambda}{3} a.\label{acc-eqn}
\end{eqnarray}
The equation (\ref{acc-eqn}) is sometimes referred as the Raychaudhuri equation \cite{Das2012, HE1973}. It provides the cosmic acceleration
which is governed by forces on the right hand side of (\ref{acc-eqn}).\\

The field equations yield

\begin{eqnarray}
\frac{1}{2} \dot{a}^2- \left(\frac{4 \pi}{3} \rho+ \frac{\Lambda}{6}\right) a^2 &=& -\frac{k_0}{2}~.\label{friedmann3}
\end{eqnarray}
The equation (\ref{friedmann3}) stands for the conservation equation of the total
energy of the universe with kinetic energy $\frac{1}{2} \dot{a}^2$, and the gravitational
potential energy as the second part on the left hand side of (\ref{friedmann3}).\\

Now, we go back to the equations (\ref{friedmann1}), (\ref{friedmann2}), (\ref{conservation}), and (\ref{friedmann3})
to obtain the general solution of the system. We define a function

\begin{eqnarray}
M (\rho)&:=& \exp\left[\int \frac{d \rho}{\rho+ p}\right] > 0,\label{function}
\end{eqnarray}
where the pressure $p= p (\rho)$. So, we get

\begin{eqnarray}
\frac{dM (\rho)}{d \rho}&=& \frac{M (\rho)}{\rho+ p}> 0.\label{differential}
\end{eqnarray}
The conservation equation (\ref{conservation}) reduces to

\begin{eqnarray}
\frac{d}{dt}\left[\ln M (\rho)+ \ln a^3\right]&=& 0,
\end{eqnarray}
or,

\begin{eqnarray}
M (\rho)\, a^3 &=& m_0,\label{restriction}
\end{eqnarray}
which is a positive constant of integration. By the inverse function theorem \cite{MT1988}, the inverse function
$M^{-1}$ exists such that

\begin{eqnarray}
\rho&=& M^{-1} (m_0 a^{-3}).\label{energy-density}
\end{eqnarray}
Let us now solve the field equations for the function $a (t)$. From (\ref{friedmann3}) and (\ref{energy-density}), we have

\begin{eqnarray}
\frac{dt}{da}&=& \Bigl[\frac{1}{3}\left(k M^{-1} (m_0 a^{-3})+\Lambda \right)a^2-k_0\Bigr]^{-1/2},\label{diffeqn1}
\end{eqnarray}
Hence,

\begin{equation}
t= \int \Bigl[\frac{1}{3}\left(k M^{-1} (m_0 a^{-3})+\Lambda\right)a^2-k_0\Bigr]^{-1/2} da + t_0,\label{solution1}
\end{equation}
where $t_0$ is the constant of integration. The right hand side of the above equation (\ref{solution1}) is a function
of the scale factor $a$. The density was earlier expressed in the equation (\ref{energy-density}).\\

In the next sections, we consider the applications of the general solution (\ref{solution1}) for a few special cases
of the equation of state.

\section{The Linear Equation of State}
\label{sec:lineos}

We start our analysis with the linear equation of state between density and pressure
is given explicitly in the form

\begin{eqnarray}
S (\rho, p)&=& c_1^2 \rho+ c_2^2 - p,\nonumber
\end{eqnarray}
or,

\begin{eqnarray}
p&=& c_1^2 \rho+ c_2^2.\label{linear-eos}
\end{eqnarray}
Here, $c_1 \neq 0$, $c_2$ are two prescribed constants. Using (\ref{function}), (\ref{energy-density}), and (\ref{linear-eos}), one
can deduce that

\begin{eqnarray}
M (\rho)&=& \left[(1+c_1^2) \rho+ c_2^2\right]^{\frac{1}{1+c_1^2}}= m_0 a^{-3}.\label{function2}
\end{eqnarray}
Hence,

\begin{eqnarray}
\rho &=& M^{-1}[(m_0 a^{-3})]= (1+c_1^2)^{-1} \left[(m_0 a^{-3})^{1+c_1^2}-c_2^2\right],\label{edensity-lin-eos}
\end{eqnarray}
and from (\ref{linear-eos}), we have

\begin{eqnarray}
p &=& c_1^2 (1+c_1^2)^{-1} \left[(m_0 a^{-3})^{1+c_1^2}-c_2^2\right]+ c_2^2.\label{pressure-lin-eos}
\end{eqnarray}
For a homogeneous linear equation of state, we have to put $c_2= 0$. Thus, we have

\begin{eqnarray}
\rho&=& (1+ c_1^2)^{-1} \left(m_0 a^{-3}\right)^{1+c_1^2},\nonumber
\end{eqnarray}
and from (\ref{solution1}) with $\Lambda= 0$, we get

\begin{eqnarray}
t &=& \int \left[\frac{k}{3} (1+c_1^2)^{-1} (m_0 a^{-3})^{1+c_1^2} a^2- k_0\right]^{-1/2} da + t_0.\label{solution1.1}
\end{eqnarray}
In the radiation era \cite{Weinberg1972}, the constant $c_1= 1/\sqrt{3}$. Thus, Eq. \ref{solution1.1} implies that

\begin{eqnarray}
t- t_0 &=& \int \frac{a da}{\sqrt{k \left(\frac{m_0^{2/3}}{2}\right)^2- k_0 a^2}},\label{solution1.2}
\end{eqnarray}
We work out explicitly on the integral (\ref{solution1.2}) above for three cases $k_0= 0,~-1~+1$ as follows:\\\\

\subsection{Flat universe ($k_0= 0$)}

For the spatially flat universe, i.e. when $k_0= 0$, using the above integral (\ref{solution1.2}), one arrives at

 \begin{eqnarray}
 t-t_0 &=& \frac{a^2}{\sqrt{k} m_0^{2/3}}> 0,\label{case1a}
 \end{eqnarray}
which can be solved to 

\begin{eqnarray}
a (t) &=& k^{\frac{1}{4}} m_0 ^{\frac{1}{3}} \sqrt{t-t_0}.\label{case1b}
\end{eqnarray}
This is a well known result for the radiation filled universe. This predicts a singularity at $t= t_0$.

\subsection{Open universe ($k_0= -1$)}

Similarly, for the open universe, the integral (\ref{solution1.2}) reduces to 

\begin{eqnarray}
 t-t_0 &=& \sqrt{k \left(\frac{m_0^{2/3}}{2}\right)^2+ a^2},\label{case2a}
\end{eqnarray}
which gives

\begin{eqnarray}
a (t) &=& \sqrt{(t-t_0)^2- k \left(\frac{m_0^{2/3}}{2}\right)^2} \,\,\,\,\,> 0. \label{case2b}
\end{eqnarray}
This is an ever expanding model of the universe. We find that the solution predicts a singularity at
$t  = t_0 \pm k \left(\frac{m_0^{2/3}}{2}\right)^2$

\subsection{Closed universe ($k_0= 1$)}

Finally,  for the closed universe, 
\begin{eqnarray}
a (t) &=& \sqrt{k \left(\frac{m_0^{2/3}}{2}\right)^2- (t- t_0)^2} > 0. \label{case3a}
\end{eqnarray}

In this case, it is easily seen that $a (t)= 0$,  when $(t- t_0)^2= k \left(\frac{m_0^{2/3}}{2}\right)^2$,
that means when $t-t_0= \pm \sqrt{k \left(\frac{m_0^{2/3}}{2}\right)^2}$, which presents two situations at extreme
points big bang and the big crunch. These are usual consequences for positive spatial curvature ($k_0 = 1$).

\section{The equation of state for a Chaplygin gas}
\label{sec:Chaplygin}

The equation of state in this case satisfies the following equation \cite{DBC2004}

\begin{eqnarray}
S (\rho, p) &=& p- \gamma \rho + \frac{A}{\rho^n},\nonumber
\end{eqnarray}
that means,  

\begin{eqnarray}
p &=& \gamma \rho- \frac{A}{\rho^n},\label{mcg1}
\end{eqnarray}
where $\gamma$, $A$ and $n$ are prescribed constants. The case for $A= 0$ has been discussed in section \ref{sec:lineos}.
The case $\gamma= 0$ represents a
polytropic equation of state for negative $n$ and $A$. The equation of state $p= p (\rho)$ mentioned in the equation (\ref{mcg1}) is interesting. Usually it is called the equation of state for a modified Chaplygin gas. 

Following the procedure given in (\ref{condition3}), we have

\begin{eqnarray}
\frac{d p}{d \rho} &=& -\left[\frac{\partial S}{\partial \rho}/\frac{\partial S}{\partial p}\right]= \gamma+ \frac{nA}{\rho^{n+1}},\label{mcg2}
\end{eqnarray}
and from (\ref{function}), we have

\begin{eqnarray}
M (\rho) &=& \exp \left[\int \frac{d \rho}{(\gamma+1) \rho- \frac{A}{\rho^n}}\right] = \Bigl[(\gamma+1) \rho^{n+1}- A\Bigr]^{1/\nu} ,\label{mcg3}
\end{eqnarray}
where $\nu= (\gamma+1)(n+1)$, and



\begin{eqnarray}
\frac{d M (\rho)}{d \rho} &=& \frac{1}{\nu} (\gamma+ 1) (n+ 1) \rho^n \left[(\gamma+1) \rho^{n+1}- A\right]\nonumber\\
&=& \frac{\rho^n M (\rho)}{(\gamma+1) \rho^{n+1}- A}= \frac{M (\rho)}{(\gamma+1) \rho- \frac{A}{\rho^n}}= \frac{M (\rho)}{p+ \rho}.\label{mcg5}
\end{eqnarray}
Now, from (\ref{restriction}) and (\ref{mcg3}), one gets


\begin{eqnarray}
\rho^{n+1} &=& \frac{1}{\gamma+1} \Bigg[A+ \left(\frac{m_0}{a^3}\right)^\nu\Bigg].\label{mcg7}
\end{eqnarray}
The pressure can be obtained from Eq. (\ref{mcg1}). Finally, the Eq.  (\ref{solution1}) leads to the relation

\begin{eqnarray}
t- t_0 &=& \int \Bigg[\frac{1}{3} \left\{k \left(\frac{1}{\gamma+1} \left(A+ \left(\frac{m_0}{a^3}\right)^\nu\right)\right)^{\frac{1}{n+1}}+ \Lambda\right\}a^2- k_0\Bigg]^{-1/2} da~.\label{mcg8}
\end{eqnarray}
Thus, the above expression yields the scale factor as a function of time --- at least in an implicit form.

\section{Nonlinear equation of state}
\label{section:nonlinear}

We consider a general equation of state of the form $p= -\rho- f(\rho)$, where $f(\rho)$ is any continuous function of the energy density. This kind of equation of state is very interesting because it can cover a wide ranges of models for different functional forms for $f(\rho)$. For $p = -\rho + \gamma \rho^{\lambda}$ (where $\gamma \neq 0$, and $\lambda$ are real constants), different inflationary solutions including the well known power law and exponential were found in \cite{Barrow:1990vx}. Later on this equation of state has been investigated in a series of papers with interesting fatures \cite{Nojiri:2004pf,Nojiri:2005sx, Nojiri:2005sr, bamba,haro}. As a specific and an academic example, we introduce a special equation of state: $p= - \rho- C \rho^{\alpha}$, where $\alpha \neq 1$, $C$ can be any real numbers, which was introduced in \cite{Nojiri:2004pf}. This equation of state can represent a quintessence type or a phantom like dark energy depending on the sign of $C$. Now, applying the same procedure, one may see that the energy density can be solved as


\begin{align}
\rho^{1-\alpha} & = C\, (\alpha - 1) \, \ln \left(\frac{m_0}{a^3}\right).
\end{align}
As $\rho$ is real and positive, so, depending on the sign of the function $\ln \left(\frac{m_0}{a^3}\right) $,
which is positive at very early time, and negative at very late time, we could have different scenarios. Thus, at early time, as $\ln \left(\frac{m_0}{a^3}\right)$ is positive, so we must have $C (\alpha-1)> 0$, that implies,
either ($C> 0,~\alpha> 1$), or ($C< 0, ~\alpha< 1$). At late time, i.e.,
when $a \longrightarrow \infty$, $\ln \left(\frac{m_0}{a^3}\right)$ is negative,
and hence in order $\rho$ to be real, we must have $C (\alpha-1) < 0$, that means
either ($C< 0,~\alpha> 1$), or ($C> 0,~\alpha< 1$).

However, for the above equation of state, the scale factor can be solved as (for $\Lambda= 0$, and $k_0= 0$)

\begin{align}\label{sp1-aasp}
a^3 & = m_0 \exp\left[-\,D\, (t- t_0)^{\frac{2 (1-\alpha)}{1-2\alpha}}\right],
\end{align}
where $t_0$ is some integration constant and $D$ is also constant given by

\begin{align}
D & = \Bigg[\sqrt{24\,\pi}\, \left( \frac{2\, \alpha -1}{2 (1-\alpha)}\right) \left(C \, (\alpha - 1)\right)^{\frac{1}{2\,(1-\alpha)}} \Bigg]^{\frac{2(1-\alpha)}{1-2\alpha}}.
\end{align}

As an example, we consider the following equation of state when $f (\rho) = B \rho^{1/2}$, i.e., $p = -\rho - B \sqrt{\rho}$ \cite{Frampton}.
Thus, in that case, we apply the similar procedure as we have done in the previous sections. Hence, we get

\begin{align}
M (\rho) & = \exp \left(-\frac{2}{B} \sqrt{\rho}\right)= m_0 a^{-3},
\end{align}
and the energy density takes the form

\begin{align}
\rho & = \frac{B^2}{4} \left(\ln \left(\frac{a^3}{m_0}\right)\right)^2.
\end{align}

The scale factor can be solved as (for $k_0=0$, $\Lambda= 0$)

\begin{align}\label{sp2-aasp}
a^3 & = m_0 \exp\Bigl[-\,\exp\left( -\, \sqrt{6 \pi}\, \left|B \right| \, \left(t-t_0\right)\right)\Bigr],
\end{align}
where $t_0$ is some constant of integration.
Thus, we find that for inhomogeneous equations of state, the
scale factor can be explicitly solved as the function of cosmic time,
and hence the other cosmological  parameters as well. However, it is readily seen that the solutions carry some important information about the evolution of the universe, for instance, in equations (\ref{sp1-aasp}), (\ref{sp2-aasp}), one can see that the scale factor does not allow any finite time singualrity, moreover it shows an exponential expansion of the universe. Finally, there is another equation of state of $f (\rho)= \rho^{\eta} - 1$ (see Ref. \cite{bamba}) (where $\eta$ being a real number) is
which is very similar to the $\Lambda$CDM cosmological model for $|\eta|\ll 1$.
By employing similar methodology for this equation of state, one may calculate the geometric variables,
such as, the scale factor, Hubble rate etc, in terms of the cosmic time.

\section{Summary and conclusions}
\label{conclu}

In the last couple of years, physical cosmology has seen its successive developments with different new theories as well as the observational data. In particular, lots of dark energy models in the context of Einstein gravity and diferent modified gravity theories, alternatives to the Einstein gravity, have been the central theme to understand the dynamical evolution of the universe. Nevertheless, the actual dynamics of the universe remains a mystery for several reasons. In Einstein gravity, one mainly focuses on the equation of state of the underlying fluid and try to solve the Einstein's equations. Following this approach, with some simple and complicated choices of $p = \psi(\rho)$, where $p$, $\rho$ are respectively the pressure, energy density of the underlying fluid connected by any anylytic function $\psi$,  the evolutions equations, mainly, the expansion scale factor is determined, and hence the dynamics of the universe. For implicit relations between $p$ and $\rho$, which could also be a justified possibility, mere attention is furnished. The current work is an attempt to establish a general way in order to solve the evolution equations for the scale factor. 
Thus, considering the FLRW universe as the background geometry, we have established a general formalism to find the expansion scale factor when the equation of state of the cosmic fluid, whether it is an explict relation between $p$ and $\rho$, or the implict, is given. We have shown how the particular cases, i.e. when the equation of state is linear, or Chaplygin type fluid, or even if the equation of state is more complicated, are recovered using the general formalism that we have provided. Thus, in summary, the current work offers a new and interesting technique to solve the evolution equations of the universe when the cosmic fluid of the background universe follows a general equation of state irrespective of any individual functional relation between $p$ and $\rho$.

\section*{Acknowledgments}
 SP acknowledges Science and Engineering Research Board (SERB), Govt. of India for National Post-Doctoral Fellowship (File No. PDF/2015/000640). Also, SP thanks the Department of Mathematics, Jadavpur University where a part of the work was carried out.

\end{document}